\documentclass [12pt]{article}
\makeatletter
\def\section{\@startsection {section}{1}{\z@}{-3.5ex plus -1ex minus
 -.2ex}{2.3ex plus .2ex}{\large\bf}}
\def\subsection{\@startsection{subsection}{2}{\z@}{-3.25ex plus -1ex 
minus -.2ex}{1.5ex plus .2ex}{\normalsize\bf}}
\makeatother
\makeatletter
\def\theequation{\arabic{section}.\arabic{equation}}
\newcommand{\sect}[1]{\setcounter{equation}{0}\section{#1}}
\@addtoreset{equation}{section}
\renewcommand{\theequation}{\thesection.\arabic{equation}}
\makeatother
\newcommand{\ky}{\kappa_{ \rm orb}}

\newcommand{\ket}[1]{|{#1}\rangle}
\newcommand{\braket}[2]{\langle{#1}|{#2}\rangle}
\newcommand{\be}{\begin{equation}}
\newcommand{\ee}{\end{equation}}
\newcommand{\bea}{\begin{eqnarray}}
\newcommand{\eea}{\end{eqnarray}}
\def\one{{\rm 1\kern -.9mm l}}

\begin{document}
\begin{titlepage}
\rightline{DFTT 20/2001}
\rightline{NORDITA-2001/15 HE}
\vskip 1.8cm
\centerline{\Large \bf ${\cal N}$ = 2 GAUGE THEORIES ON SYSTEMS}
\vskip 0.4cm
\centerline{\Large \bf OF FRACTIONAL D3/D7 BRANES 
\footnote{Work partially supported by the European Commission RTN
programme HPRN-CT-2000-00131 and by MURST.}}
\vskip 1.4cm 
\centerline{\bf M. Bertolini $^a$, P. Di Vecchia $^a$,
M. Frau $^b$, A. Lerda $^{c,b}$, 
R. Marotta $^{a,d}$}
\vskip .8cm \centerline{\sl $^a$ NORDITA, Blegdamsvej 17, DK-2100
Copenhagen \O, Denmark}
\vskip .4cm \centerline{\sl $^b$ Dipartimento di  Fisica Teorica,
Universit\`a di Torino} \centerline{\sl and I.N.F.N., Sezione di
Torino,  Via P. Giuria 1, I-10125 Torino, Italy}
\vskip .4cm \centerline{\sl $^c$ Dipartimento di Scienze e Tecnologie
Avanzate} \centerline{\sl Universit\`a del Piemonte Orientale, I-15100
Alessandria, Italy}
\vskip .4cm \centerline{\sl $^d$ Dipartimento di Scienze Fisiche, 
Universit\`a di Napoli} 
\centerline{\sl Complesso Universitario Monte S. Angelo, Via Cintia, I-80126
Napoli, Italy}
\vskip 1.8cm
\begin{abstract}
We study a bound state of fractional D3/D7-branes in the ten-dimensional space 
${\rm I\!R}^{1,5}\,\times\,{\rm I\!R}^{4}/{\bf Z_2}$ using the boundary 
state formalism. We construct the boundary actions for this system and 
show that higher order terms in the twisted fields are needed in order to 
satisfy the zero-force condition. We then find the classical background 
associated to the bound state and show that the gauge theory living on a 
probe fractional D3-brane correctly reproduces the perturbative behavior of
a four-dimensional ${\cal{N}} =2$ supersymmetric gauge theory with fundamental
matter. 
\end{abstract}
\end{titlepage}
\renewcommand{\thefootnote}{\arabic{footnote}}
\setcounter{footnote}{0} \setcounter{page}{1} 
\tableofcontents  
\vskip 1cm
\section{Introduction}
\label{intro}
During the last few years, it has become more and more evident that
the low-energy properties of D-branes can be studied in two complementary
ways: one based on the fact that a D-brane is a classical solution
of the string effective theory (supergravity) charged under
a RR potential; the other based on the fact that
a D-brane supports a gauge field theory on its
world-volume. The realization of this two-fold interpretation, which in 
the literature goes under the name of gauge/gravity 
correspondence, has been one of the most significant results of 
the recent research in string theory. In fact, in view of this correspondence,
one can exploit the classical geometrical 
properties of D-branes to study gauge theories
or, vice-versa, use the quantum properties of the gauge theory defined
on a D-brane to study the dynamics of non-perturbative
extended objects.

In the case of the D3-branes of the type IIB string theory in flat space, 
it was possible to carry this correspondence much further by taking the 
so-called near-horizon limit, 
and observing that in this limit the gravity degrees of freedom
(closed strings) propagating in the entire ten-dimensional space decouple from
the gauge degrees of freedom (open strings) living on the four-dimensional 
world-volume of the D3-brane. This decoupling led Maldacena~\cite{mal} to 
conjecture an exact duality between the ${\cal{N}}=4$ super Yang-Mills 
theory in four dimensions, which is the conformal field theory
living on the D3-brane world-volume, and type IIB string theory on 
$AdS_5 \times S_5$, which is the space-time geometry 
in the near horizon limit. This remarkable conjecture, which has been confirmed
by all subsequent studies, has opened the way to the use of brane dynamics in 
the analysis of the strong coupling regime of four-dimensional gauge 
theories. 

More recently, a lot of efforts have been made to find possible
extensions of the Maldacena duality to less supersymmetric and non-conformal
gauge theories with more realistic properties such as asymptotic freedom and 
a running coupling constant. 
The simplest theories with these features are those with ${\cal{N}}=2$ 
supersymmetry that can be obtained, for instance, by studying fractional 
branes on orbifolds~\cite{Douglas:1996sw,Johnson:1997py,Douglas:1997xg,Diaconescu:1998br}. 
In particular, the world-volume theory defined on a stack of 
fractional D3-branes in ${\rm I\!R}^{1,5}\times{\rm I\!R}^{4}/{\bf Z_2}$ 
is a pure ${\cal N}=2$ super Yang-Mills theory in 
four dimensions~\cite{Klebanov:2000rd},
and thus this is a very natural system to consider in order to
study possible non-conformal 
extensions of the Maldacena duality.
However, as shown in Ref.s~\cite{Bertolini:2000dk,Polchinski:2000mx} 
(and for more general orbifolds in Ref.~\cite{marco}) the supergravity 
solutions corresponding to these fractional D3-branes possess naked 
singularities of repulson type \cite{kal}. 
The appearance of naked singularities is a quite general 
feature of the supergravity solutions that are dual to non-conformal
gauge theories, but in the orbifold case, it seems that there exists 
a general mechanism to resolve them.
Indeed, by analyzing the action of a probe D3-brane 
in the singular background, one can see that
the probe becomes tensionless before reaching the repulson singularity
on a hypersurface called enhan\c{c}on~\cite{enhanc} (similar results hold 
also in the case of fractional branes on compact orbifolds, see 
Ref.~\cite{anto}). This fact suggests that at the enhan{\c{c}}on the 
classical solution cannot be trusted anymore because additional light 
degrees of freedom come into play and the supergravity 
approximation is no longer correct. 
Therefore, the presence of the enhan\c{c}on allows
to consistently excise the singularity region and obtain a well-behaved 
solution~\cite{excision}, but at the same time
it does not allow to easily take the decoupling limit anymore
(for a discussion on the physics of the enhan\c{c}on for these and more 
complicated systems see Ref.s~\cite{Aharony,Petrini:2001fk}).

Despite these problems, the classical solution describing fractional 
D3-branes on orbifolds has been successfully 
used to study the perturbative 
dynamics of ${\cal{N}}=2$ supersymmetric gauge theories and obtain 
their correct perturbative 
moduli space~\cite{Bertolini:2000dk,Polchinski:2000mx,marco}. 
Among other things, this analysis has also shown
that the enhan{\c{c}}on corresponds, in the gauge theory, 
to the scale where the gauge 
coupling constant diverges (the analogue of $\Lambda_{\rm QCD}$ in QCD). 
These results, that
seem to be in contrast 
with a duality interpretation \`{a} la Maldacena where the supergravity 
solution gives a good description of the gauge theory for {\it large} 't 
Hooft coupling, can instead be easily understood if we regard the 
classical supergravity solution as an effective way of summing over 
all open string loops, as explained for example in Ref.~\cite{nbps}.
{F}rom this point of view, in fact, one does not take the near-horizon limit 
({\it i.e.} $r \rightarrow 0$, where $r$ is the distance from the 
source branes), but rather expands the classical solution
around $ r \rightarrow \infty$ 
where the metric is almost flat and the 
supergravity approximation is valid. This expansion corresponds
to summing closed string diagrams at tree level, but, because of 
the open/closed string duality, it is also equivalent to summing
over open string loops. Therefore, expanding the supergravity
solution around $r\to \infty$ is equivalent to perform an expansion
for {\it small} 't Hooft coupling.
In view of these considerations, it is then not surprising that the 
supergravity solution of Ref.s~\cite{Bertolini:2000dk,Polchinski:2000mx,marco}
encodes the perturbative properties of the ${\cal{N}}=2$
gauge theory living on the world-volume of a fractional D3-brane, 
but at the same time it is also natural to expect that this approach
does not include 
the non-perturbative instanton corrections to the moduli space.
In conclusion we can say that the previous results are not a 
consequence of a Maldacena-like duality, but rather they follow 
directly from the gauge/gravity correspondence or, equivalently, 
from the open/closed 
string duality~\footnote{Recent papers stressing the importance of the 
open/closed string duality in our context 
are in Ref.s~\cite{Bianchi,Kaku,park}.}. 
On the other hand, the presence of the enhan\c{c}on, which excises 
the region of space-time corresponding to scales where non-perturbative 
effects become  relevant in the gauge theory and which is reminiscent 
of the curve separating strong from weak coupling in ${\cal N}=2$ super 
Yang-Mills theory \cite{fayya}, is consistent with the above picture.
To incorporate in this scenario also the non-perturbative instanton 
effects, presumably one must include D-instanton corrections already 
at the string level~\cite{GreenGut,GreenGut2}, as done in Ref.~\cite{TRICOLORI}
for the ${\cal{N}}=4 $ super Yang-Mills theory, or start directly from
M-theory~\cite{GGV}. M-theory is also the starting 
point of a very interesting alternative approach pursued recently in 
Ref.s~\cite{ansar,Brinne:2000fh} where the non-perturbative properties 
of ${\cal{N}} =2$ supersymmetric gauge theories have been obtained
by taking the near-horizon limit. More recently, the instanton 
corrections for systems of fractional D-branes have been discussed 
in Ref. \cite{fu},
while alternative approaches to investigate 
supersymmetric ${\cal{N}}=2$ gauge theories, at the perturbative level,
have been carried out for instance in Ref.s~\cite{Petrini,Pol,dario,zaffa}.

In a recent paper~\cite{GRANA2}, the approach of 
Ref.s~\cite{Bertolini:2000dk,Polchinski:2000mx} has been extended to
the case of a system of fractional D3/D7-branes on the 
orbifold ${\rm I\!R}^{1,5}\times{\rm I\!R}^{4}/{\bf Z_2}$ 
and the perturbative properties of 
the ${\cal{N}}=2$ supersymmetric gauge theory living on the
D3-brane world-volume have again been recovered from supergravity. 
In this case the gauge theory
includes also hypermultiplets in the fundamental representation,
associated to the open strings stretched between the
D3 and the D7-branes. 

In this paper, motivated by the considerations discussed above 
and using essentially 
the information provided by the boundary state formalism, we discuss further 
properties of the fractional D7-branes and of the bound states of
fractional D3/D7-branes 
in the orbifold background.
In particular we construct the boundary action for a fractional 
D7-brane 
and find that in order to satisfy the no-force condition required for
a supersymmetric system, terms of higher order in the
twisted fields, which are not accounted by the boundary
state, must be included.
We then solve explicitly the supergravity field equations 
for the D3/D7 system and study the properties of the dual four-dimensional 
gauge theory, finding agreement with the analysis of Ref.~\cite{GRANA2}. 
An interesting feature of this system is that, unlike the 
case discussed in 
Ref.s~\cite{Bertolini:2000dk,Polchinski:2000mx,marco}, the twisted fields
receive contribution from diagrams with an arbitrary number of open string 
loops or, equivalently, from closed string tree diagrams containing an 
arbitrary number of boundaries. However, as expected from the 
${\cal{N}}=2$ non-renormalization theorems, we find 
that in the gauge theory
the twisted fields appear always in special combinations in which only 
the one-loop perturbative contribution is non-trivial. 

The paper is organized as follows. In Sect.~\ref{section2} we write 
the field equations of type IIB supergravity on the orbifold
coupled to fractional branes. In Sect.~\ref{section3} 
we use the boundary state to study 
the properties of the fractional D7-branes determining their couplings 
with the bulk fields and the first order approximation 
of their classical solution. 
In Sect.~\ref{section4} 
we extend the previous analysis to a bound state of fractional 
D3/D7-branes finding its complete classical solution. Using the no-force 
argument, we are also able to fix 
the complete form of the D7 brane boundary action.
Finally in Sect.~\ref{section5}, by means of the probe analysis, we derive the 
perturbative behavior of the corresponding ${\cal{N}}=2$ 
gauge theory and discuss its properties. 
A few technical results on the boundary state construction are 
reviewed in Appendix A.

\vskip 1cm
\sect{Field equations for fractional D-branes}
\label{section2}
We consider type IIB supergravity in ten dimensions on the 
orbifold 
\be 
{\rm I\!R}^{1,5}\,\times\,{\rm I\!R}^{4}/{\bf Z_2}
\label{orbifold}
\ee 
where ${\bf Z_2}$ is the reflection parity along
$x^6$, $x^7$, $x^8$ and $x^9$. Its action (in the Einstein frame) can be
written as~\footnote{Our conventions for curved indices and forms are the
following:
$\varepsilon^{0\dots9}=+1$; signature $(-,+^9)$;
$\mu,\nu=0,\dots,9$; $\alpha,\beta=0,\ldots,3$; $i,j=4,5$;
$\ell,m=6,\ldots,9$; $\omega_{(n)}={1\over n!}
\,\omega_{\mu_1 \dots \mu_n} dx^{\mu_1}\wedge\dots\wedge  dx^{\mu_n}$, and
$*\omega_{(n)}={\sqrt{-\det G}\over n!\,(10-n)!}\,
\varepsilon_{\nu_1\dots\nu_{10-n}\mu_1 \dots \mu_n} \,\omega^{\mu_1 \dots
\mu_n}  dx^{\nu_1}\wedge\dots\wedge dx^{\nu_{10-n}}.$}
\[
S_{\rm IIB} = \frac{1}{2 \ky^2} \Bigg\{ \int d^{10} x~
 \sqrt{-\det G}~ R - \frac{1}{2} \int \Big[ d \phi \wedge {}^* d \phi
 \,+\, {\rm e}^{- \phi} H_{3}  \wedge {}^* H_{3}\,+\, {\rm e}^{2
 \phi}\, F_{1} \wedge {}^* F_{1}
\]
\begin{equation}
 + \,\,{\rm e}^{\phi} \,{\widetilde{F}}_{3} \wedge {}^*
 {\widetilde{F}}_{3} \,+\, \frac{1}{2}\, {\widetilde{F}}_{5}
 \wedge {}^* {\widetilde{F}}_{5}  \, -\,  C_{4} \wedge H_{3}
 \wedge F_{3} \Big] \Bigg\}
\label{tendim3}
\end{equation}
where 
\be 
H_{3} = d B_{2}~~~,~~~F_{1}=d C_{0}~~~,~~~ F_{3} =
d C_{2}~~~,~~~F_{5} = d C_{4}
\label{form2}
\ee
are, respectively, the field strengths of the NS-NS
2-form and the 0-, 2- and 4-form
potentials of the R-R sector, 
and~\footnote{Notice that we use definitions where 
$B_2 \rightarrow - B_2$ with respect to Ref.~\cite{Bertolini:2000dk}.}
\be 
{\widetilde{F}}_{3} = F_{3}
+ C_{0} \wedge H_{3}~~~~ ,~~~~{\widetilde{F}}_{5} = F_{5} +
C_{2} \wedge H_{3}~~.
\label{form3}
\ee 
Moreover, $\ky^2\equiv (2\pi)^{7}\,g_s^2\,\alpha'^4=2\kappa^2$ where $g_s$
is the string coupling  constant, and the self-duality constraint
${}^* {\widetilde{F}}_{5}={\widetilde{F}}_{5}$  has to be
implemented on shell. 

We are interested in obtaining classical solutions of the field equations
descending from the action (\ref{tendim3}) that describe fractional D branes
characterized by the presence of ``twisted'' scalar fields
$b$ and $c$ defined as 
\begin{equation}
C_{2} = c~ \omega_2~~~~~,~~~~B_{2} = b~ \omega_2
\label{wra98}
\end{equation}
where $\omega_2$ is the anti-self dual 2-form
associated to the vanishing 2-cycle of the orbifold ALE space. In
our normalizations
\cite{Bertolini:2000dk}, this form satisfies
\begin{equation}
\int_{\rm ALE} \omega_2 \wedge \omega_2 = -1~~.
\label{omega2co}
\end{equation}
Inserting eq.(\ref{wra98}) in the action (\ref{tendim3}), we easily get  
\[
S'_{\rm IIB} =  \frac{1}{2 \ky^2} \Bigg\{ \!\int\! d^{10}x\, \sqrt{- \det
G}\, R - \frac{1}{2} \int\! \left[ d \phi \wedge {}^* d \phi + 
e^{2 \phi} d C_0 \wedge {}^* d C_0 +\frac{1}{2}  
 {\widetilde{F}}_{5} \wedge {}^*
{\widetilde{F}}_{5}  \right]
\]
\begin{equation}
- \frac{1}{2} \int  \left[ e^{-\phi} \,d b  \wedge {}^{*_6} d b + 
e^{\phi} ( d c + C_0 \,db) \wedge {}^{*_6} (dc + C_0 \,db) +
C_4  \wedge d b \wedge d c  \,\right]_6
\Bigg\}
\label{sred5}
\end{equation}
where the index $6$ refers to the six-dimensional space orthogonal 
to the orbifold directions.

The field equations for a fractional D-brane are obtained by varying the
total action
\be
S'_{\rm IIB} + S_{\rm b}
\label{stot}
\ee
where $S_{\rm b}$ describes the coupling of the bulk supergravity fields
with the brane; we recall however, that 
only the linear part of the boundary action $S_{\rm b}$ is relevant to yield
the source terms for the inhomogeneous field equations \cite{nbps}.
For the moment we do not specify the form of $S_{\rm b}$ which instead will be
discussed in the following sections for the specific cases of the fractional D3
and D7 branes. Defining
\be
\Omega_4 \equiv \delta(x^6)\cdots \delta(x^9)\,dx^6\wedge \cdots
\wedge dx^9~~,
\label{omega4}
\ee
the field equation for the dilaton is
\bea
&&\!\!\!\!\!\!\!\!\!\!\!\!\!
d {}^{*} d \phi - e^{2 \phi} d C_{0} \wedge {}^{*} d C_{0} 
+ \frac{1}{2} \left[ e^{- \phi} db \wedge {}^{*_6} d b 
\right. \label{dila12} \\
&&~~~~\left.
- e^{\phi} 
(d c + C_{0}\, db ) \wedge {}^{*_6} (dc + C_{0} \,db)  
\right]\wedge \Omega_4
+ 2 \ky^{2} \, \frac{\delta S_{\rm b}}{\delta \phi}
=0~~,
\nonumber 
\eea
and the one for the axion is
\begin{equation}
d\, ( {\rm e}^{2 \phi}\,{}^{*}d C_{0} ) - {\rm e}^{\phi} db \wedge {}^{*_6}
(dc +C_{0} \,db) \wedge \Omega_4 + 2 \ky^{2}\,
\frac{\delta S_{\rm b}}{\delta C_{0}} 
=0 ~~.
\label{axion45}
\end{equation}
The field equations for the two twisted fields $b$ and $c$ are 
respectively
\begin{equation}
d \left[ {\rm e}^{- \phi} \,\, {}^{*_6}db 
+ C_{0} \,{\rm e}^{\phi}\,\,{}^{*_6} 
( dc + C_{0} \,db )\right]
+ {\widetilde{F}}_{5} 
\wedge dc + 2 \ky^{2} \frac{\delta S_{\rm b}}{\delta b}=0
~~,
\label{beq76}
\end{equation}
and 
\begin{equation}
d \left[ {\rm e}^{\phi}\,\, {}^{*_6} (d c + C_{0} \,db)\right] - 
{\widetilde{F}}_{5} 
\wedge db + 
2 \ky^{2} \,\frac{\delta S_{\rm b}}{\delta c} =0~~.
\label{ceq}
\end{equation}
Finally, the field equation for the untwisted $4$-form $C_{4}$
is
\begin{equation}
d \,{}^{*} {\tilde{F}}_5 - db \wedge dc \wedge \Omega_4 + 
2 \ky^{2} \,\frac{\delta S_{\rm b}}{\delta C_{4}}=0~~,
\label{eqc4}
\end{equation}
and the one for the metric tensor is
\be
R_{\mu \nu} 
-\frac{1}{4 \cdot 4!} ({\tilde{F}}_{5})_{\mu \rho \sigma \tau 
\delta}({\tilde{F}}_{5})_{\nu}^{~\rho \sigma \tau \delta} +
2 \ky^{2} 
\frac{\delta S_b}{\delta G^{\mu \nu}}=
\frac{1}{2} \left[\partial_{\mu} \phi \,\partial_{\nu} \phi + {\rm e}^{2 \phi}
\partial_{\mu} C_{0} \,\partial_{\nu} C_{0} \right] 
+ T_{\mu\nu}^{(b,c)} 
\label{meteq23}
\ee
where $T_{\mu\nu}^{(b,c)}$ is the stress energy tensor of the scalars 
$b$ and $c$ whose explicit expression is not really needed in the following.

To analyze these equations it is convenient to introduce the following complex 
quantities~\footnote{The relation
between $G_{1}$ and the usual complex $3$-form of type IIB supergravity 
is  $G_{3} \equiv  F_{3} + \tau H_{3} = 
G_{1} \wedge \omega_2$.}
\begin{equation}
\tau = C_0 + {\rm i}\, {\rm e}^{-\phi} ~~~;~~~ 
\gamma = c + \tau \,b ~~~;~~~ G_{1} = dc + \tau \, db~~.
\label{def463}
\end{equation}
In fact, with simple manipulations 
the four equations (\ref{dila12})--(\ref{ceq}) 
can be combined into two complex differential equations for
$\tau$ and $\gamma$, which are
\be
d {}^{*} d \tau + {\rm i}\, e^{\phi}\, d \tau \wedge {}^* d \tau 
+ \frac{\rm i}{2}\, 
G_{1} \wedge {}^{*_6} G_{1}
\wedge \Omega_4 -2\,{\rm i}\,\ky^{2} \left[{\rm e}^{-\phi}
\frac{\delta S_{\rm b}}{\delta \phi} + {\rm i}\,
{\rm e}^{-2 \phi} \frac{\delta S_{\rm b}}{\delta C_0} \right] = 0 ~~,
\label{eqtau}
\ee 
and
\be
d {}^{*_6} d \gamma + {\rm i}\, {\rm e}^{\phi}\, d \tau  
\wedge {}^{*_6} G_{1}
- b \, d \,{}^{*6} d\tau  + {\rm i}\, {\widetilde{F}}_{5} \wedge G_{1}
+ 2 \,{\rm i} \,\ky^{2} \left[ \frac{\delta S_{\rm b}}{\delta b} -  
\tau \frac{\delta S_{\rm b}}{\delta c} \right] = 0 
~~.
\label{eqgamma}
\ee
In the following we are going to solve these equations for bound states 
made of fractional D3 and D7 branes. In particular
we will consider configurations in which the D7 branes extend in the
directions $x^0,..., x^3,x^6,... x^9$ 
({\it i.e.} entirely along the orbifold) and
the D3 branes in the directions $x^0,..., x^3$ ({\it i.e.} transversely to
the orbifold). With this arrangement, the twisted fields $b$ and $c$,
which are stuck at the orbifold fixed point, are functions
only of the transverse coordinates $x^4$ and $x^5$. Moreover, since the
D3 branes emit neither the dilaton $\phi$ nor the axion $C_0$,
these two fields are produced only by the D7 branes and thus they too are
functions only of the transverse coordinates $x^4$ and $x^5$.
For the remaining fields, the metric $G_{\mu\nu}$ and the self-dual
field strength ${\widetilde F}_5$, we take the standard
Ansatz for a D7/D3 system \cite{Aharony:1998xz}, namely
\begin{eqnarray}
ds^2 &=& H^{-1/2}\, \eta_{\alpha\beta}\,d x^\alpha dx^\beta
+ H^{1/2} \,\left(\delta_{\ell m}\,dx^\ell dx^m + {\rm e}^{-\phi} \delta_{ij}
 dx^i dx^j\right)  ~~,
\label{met48} \\
{\widetilde{F}}_{(5)} &=&  d
\left(H^{-1} \, dx^0 \wedge \dots \wedge dx^3 \right)+ {}^* d
\left(H^{-1} \, dx^0 \wedge \dots \wedge dx^3 \right) ~~,
\label{f5ans}
\end{eqnarray}
where the warp factor $H$ is a function of all coordinates that
are transverse to the D3 brane ($x^4,\ldots x^9$).

A drastic simplification occurs by analyzing the supersymmetry transformation 
rules of the gravitinos and dilatinos and asking for 
the solution to be supersymmetric. 
Plugging the Ansatz (\ref{met48}) and (\ref{f5ans}) in the variations of the 
gravitinos and dilatinos, one can show~\cite{GRANA2} 
that the existence of a Killing spinor implies that both $G_{1}$ 
and $\tau$ are holomorphic functions of 
$z\equiv x^4+{\rm i}\,x^5$, {\it i.e.} \footnote{The case of 
constant $\tau$ was discussed in Ref.s~\cite{GRANA,GUB} 
for the case of singular spaces 
and in Ref.~\cite{Bertolini:2001ma} for the blown-up case, 
while that for vanishing 
G-flux was discussed in Ref.~\cite{ke}. This kind of type 
IIB supersymmetric solutions 
are dual to those discussed in Ref.~\cite{Becker}.}
\begin{equation}
\partial_{{\bar{z}}} G_{1} = \partial_{\bar{z}} \tau = 0 ~~.
\label{gholo}
\end{equation}
The analyticity of $G_{1}$ and $\tau$ in turn implies that $\gamma$ 
is analytic too (see eq.(\ref{def463})); thus
the equations for $\tau$ and $\gamma$ drastically simplify and reduce to
\begin{equation}
d\, {}^{*}d \tau - 2\,{\rm i}\, \ky^{2} \left[
{\rm e}^{-\phi} \frac{\delta S_{\rm b}}{\delta \phi} + {\rm i}\,
{\rm e}^{-2 \phi} \frac{\delta S_{\rm b}}{\delta C_0}  \right]  =0~~,
\label{dilaeq52}
\end{equation}
and
\begin{equation}
d \, {}^{*_6} d \gamma + {\rm i} {\widetilde{F}}_5 \wedge G_{1} 
-  b\, d \, {}^{*_6} d \tau + 2\,
{\rm i}\, \ky^{2}
\left[ \frac{\delta S_{\rm b}}{\delta b} -  \tau\, 
\frac{\delta S_b}{\delta c} \right]
 =0~~.
\label{gammaeq28}
\end{equation}
Finally, we can see
that the field equations for the metric and $C_4$ are satisfied 
provided that $H$ be a solution of the following equation
\begin{equation}
\left(\delta^{ij}\partial_i \partial_j H + {\rm e}^{-\phi} \,
\delta^{\ell m}\partial_\ell \partial_m H\right) V_6 
- db \wedge dc \wedge \Omega_4+ 
2 \ky^{2}
\frac{\delta S_b}{\delta C_4} =0
\label{Heq19}
\end{equation}
where $V_6$ denotes the six-dimensional volume form $dx^4\wedge\cdots
\wedge dx^9$.

As anticipated, we want to find the solution of these field equations
for bound states of fractional D7/D3 branes. However, before doing this, 
we present a discussion of the pure D7 branes
on orbifold from a string theory point of view, using the boundary state
formalism.

\vskip 1cm
\section{The fractional D7-branes}
\label{section3}

In this section we will analyze in some detail the D7 branes of type IIB
in the orbifold background (\ref{orbifold}). In order to realize a BPS brane
of type IIB, 
the number $s$ of orbifold directions along its world-volume 
must be even \cite{gabstef}. Thus, for a 7 brane we have only two
possibilities: $s=2$ and $s=4$. Here we discuss only the case $s=4$,
{\it i.e.} when the brane extends throughout the entire orbifold, since this
is the relevant case to yield four-dimensional ${\cal N}=2$
gauge theories in the presence of fractional D3 branes.
 
The fact that D7 branes with $s=4$ extend entirely along the orbifold has
some peculiar consequences that we would like to emphasize. From a string 
theory point of view, these branes are sources not only 
of those fields that are typical
of a D7 brane ({\it i.e.}
the metric $G_{\mu\nu}$, the dilaton $\phi$ and the 8-form R-R potential
$C_{8}$), but also of twisted fields, which, specifically, comprise
a scalar ${\widetilde b}$ from the
twisted NS-NS sector and a 4-form potential $A_{4}$ from the
twisted R-R sector. It is interesting to observe
that these twisted fields are the same as those emitted by the fractional
D3 branes studied for example in 
Ref.s~\cite{Bertolini:2000dk,Polchinski:2000mx}.
Moreover, the charge of these D7 branes under $C_{8}$ 
is a half of that carried by the D7 branes in flat space. Thus, it is
natural to regard these configurations as fractional branes, despite the fact
that, contrarily to what happens for the fractional branes of
lower dimension, they cannot be interpreted as wrapped branes.
Finally, the charge of these D7 branes under the twisted 4-form potential
is a quarter of that carried by the fractional D3 branes.

All these features can be clearly seen by computing
the vacuum energy $Z$ of the open strings
stretched between two such D7 branes that is given by
\begin{equation}
Z = \int_0^\infty \frac{d s}{ s} ~{\rm Tr}_{\rm NS-R} \left[P_{\rm GSO}
\left(\frac{\one+g}{2} \right)  {\rm e}^{-
2 \pi s (L_0-a)} \right]
\label{part45}
\end{equation}
where $P_{\rm GSO}$ is the GSO projection, $g$ is the orbifold ${\bf Z}_2$
parity, and $a=1/2$ in the NS sector and $a=0$ in
the R sector.
When one takes the $\one$ inside the bracket, one gets half of
the contribution of the open strings stretched between two D7 branes 
in flat space, whereas when one
takes the $g$ inside the bracket one obtains 1/16 times the contribution
of the twisted sectors of the fractional D3 branes of 
Ref.s~\cite{Bertolini:2000dk,Polchinski:2000mx} (see 
Appendix A for more details). 
After performing the modular transformation $s\to 1/s$ 
and factorizing the resulting expression in the closed string channel,
one can derive the boundary state $|{\rm D}7\rangle$ associated to the D7
brane along the orbifold (for a review of the boundary
state formalism and its applications see, for example,
Ref.~\cite{antonella}; for an analysis of the boundary state
in orbifold theories see, for example, Ref.s~\cite{diacgom,gabstef,marco1}). 
The explicit expression of $|{\rm D}7\rangle$ and a discussion of its 
properties can be found in appendix A. Here we just mention that this boundary
state contains both an untwisted and a twisted part:
\be
|{\rm D}7\rangle = |{\rm D}7\rangle^{\rm U}+|{\rm D}7\rangle^{\rm T}
\label{bsd7}
\ee
The untwisted part $|{\rm D}7\rangle^{\rm U}$ is the same as that of the 
D7 branes in flat space but with a normalization differing by a 
factor of $1/\sqrt{2}\,$; the twisted part $|{\rm D}7\rangle^{\rm T}$ is, 
instead, similar to that of the fractional D3 branes 
but with a normalization differing
by a factor of 1/4. By saturating the boundary state $|{\rm D}7\rangle$ with 
the massless closed string states of the various sectors, one can determine 
which are the
fields that couple to the fractional D7 brane. In particular, following
the procedure found in Ref.~\cite{bs} and reviewed in Ref.~\cite{antonella},
one can find that in the untwisted sectors the D7 brane
emits the graviton $h_{\mu\nu}$\footnote{We recall that the graviton 
field and the metric are related by 
$G_{\mu\nu}=\eta_{\mu\nu} + 2\ky\,h_{\mu\nu}$.}, 
the dilaton $\phi$ and the 8-form potential $C_{8}$. 
The couplings of these fields 
with the boundary state are explicitly given by~\cite{anto} 
\bea 
\braket{{\rm D}7}{h} &=&
-\frac{T_7}{\sqrt{2}}\,\, h_{a}^{\,\,\, a} \,V_8~~,\nonumber \\
\braket{{\rm D}7}{\phi} &=&
-\,\frac{T_7}{\sqrt{2}\ky}\,\, \phi\,V_8~~,\nonumber \\
\braket{{\rm D}7}{ C_{8}} &=& 
\frac{T_7}{\sqrt{2}\ky} \,C_{01\ldots67}\, V_8
\label{untw9}
\eea
where $T_p =\sqrt{\pi} \,(2\pi\sqrt{\alpha '})^{(3-p)}$, appearing
in the normalization of the boundary state, is related to the brane
tension in units of the gravitational coupling constant \cite{bs,bs1},
$V_8$ is the (infinite) world-volume  of the D7 brane, and the index
$a$ labels its eight longitudinal directions. Notice that
there is no coupling of the boundary state $|{\rm D}7\rangle$
with the untwisted 4-form $C_4$, in agreement
with the observation \cite{GRANA2} that the D7 branes do not
carry charge under $C_4$.
  
By doing this same analysis in the twisted sectors, we find that, 
as advertised before, the boundary state $|{\rm D}7\rangle$ emits a massless 
scalar $\widetilde b$ from the NS-NS sector and a 
$4$-form potential $A_{4}$ from the R-R
sector. These fields exist only at the orbifold fixed hyperplane 
$x^6=x^7=x^8=x^9=0$,  and their couplings with the boundary state turn out 
to be given by~\cite{anto} 
\bea
\braket{{\rm D}7}{\widetilde b} &=& \frac{T_3}{4\sqrt{2}\,\ky}\,
\frac{1}{2\pi^2{\alpha'}}
\,\widetilde b \,\, V_4 ~~,\nonumber \\ 
\braket{{\rm D}7}{A_{4}} &=& -\,\frac{T_3}{4\sqrt{2}\,\ky}\,
\frac{1}{2\pi^2{\alpha'}}\,A_{0123}\, V_4
\label{twi86}
\eea
where $V_4$ is the (infinite) world-volume of the 7 brane
that lies outside the orbifold.

{F}rom the explicit couplings (\ref{untw9}) and (\ref{twi86}), it is
possible to infer the form of the world-volume action of a fractional
D7 brane. Of course, the boundary state approach allows to obtain only
the terms of the world-volume action that are linear in the bulk fields.
However, terms of higher order can be determined with other methods. 
For example, from our previous considerations, 
it is natural to write for the untwisted fields the same 
world-volume action of the D7 branes in flat space but with an extra overall
factor of $1/\sqrt{2}$ as dictated by the boundary state.
Therefore, we write (in the Einstein frame)
\be
\left.S_{~\rm b}^{\rm D7}\right|_{\rm U} = 
-\,\tau_7
\, \int d^8x~{\rm e}^\phi\,\sqrt{-\det
g_{ab}} ~+~
\tau_7\, \int
C_{8}
\label{sbuntw}
\ee
where $\tau_p\equiv T_p/({\sqrt{2}\,\ky})$ and $g_{ab}$ is the induced metric.
It is easy to 
check that this action correctly accounts for the couplings (\ref{untw9}).
Furthermore, recalling that $\ky=\sqrt{2}\,\kappa$, we can see
that the $C_8$ charge
of our D7 brane is a half of that carried by the D7 branes in flat space. 

For the twisted fields, instead, things are slightly more complicated. Using
the couplings (\ref{twi86}) and defining
${\widetilde \tau}_3\equiv T_3/(2\sqrt{2}\pi^2{\alpha'}\ky)$, one can write 
\be
\left.S_{~\rm b}^{\rm D7}\right|_{\rm T}
\simeq \,\frac{{\widetilde \tau}_3}{4}
\,\int d^4x~\sqrt{-\det
g_{\alpha\beta}}~{\widetilde b}\,-\,
\frac{{\widetilde \tau}_3}{4}\,\int A_4+...
\label{sbtw}
\ee
where in the first term the four-dimensional induced metric has been inserted 
to enforce
reparametrization invariance on the world-volume, and the ellipses stand
for terms of higher order which are not accounted by the boundary state
approach but which, in principle, can be present. 
In the following section we will show that such higher order terms are
indeed present in the complete world-volume action of the fractional
D7 branes. 

As explained in Ref.~\cite{bs}, the boundary state formalism allows
also to compute the asymptotic  behavior of the
various fields  in the classical brane solution.  
Applying this technique to the case of a stack of N 
coincident fractional D7 branes, we find 
that, to leading order in $Ng_s$, the metric is 
\be 
ds^2 \simeq
\eta_{ab} \,dx^b
dx^b + \left(1-\frac{Ng_s}{2\pi}\,\log\frac{\rho}{\epsilon}\right)\,
\delta_{ij}
\,dx^idx^j\,+...
\label{met1}
\ee 
where $\rho = \sqrt{(x^4)^2 + (x^5)^2}$ and $\epsilon$
is a regulator, while the dilaton is
\be 
\phi \simeq \frac{Ng_s}{2\pi}\,\log\frac{\rho}{\epsilon}
+...~~,
\label{dila}
\ee 
the 8-form R-R potential is
\be 
C_{8} \simeq \frac{Ng_s}{2\pi}\,\log\frac{\rho}{\epsilon}
~dx^0\wedge \cdots \wedge dx^7 +...~~.
\label{c41}
\ee  
and the 4-form potential $C_4$ vanishes at linear order.
The asymptotic behavior of the twisted fields is instead given by
\begin{eqnarray}
{\widetilde b} &\simeq& -\,Ng_s\pi\alpha'\,\log\frac{\rho}{\epsilon}
+...~~,
\label{b1} \\
A_{4} &\simeq& -\,Ng_s\pi\alpha'\,\log\frac{\rho}{\epsilon}
~dx^0\wedge \cdots \wedge dx^3 +...~~.
\label{a41}
\end{eqnarray}
If we insert these expressions into the world-volume action
\be
S_{~\rm b}^{\rm D7} =
\left.S_{~\rm b}^{\rm D7}\right|_{\rm U}+ 
\left.S_{~\rm b}^{\rm D7}\right|_{\rm T}
\label{actionbound}
\ee
and consistently retain only terms of first order in $Ng_s$,
we obtain a constant result, thus verifying the no-force condition
at first order. 

To extend this analysis to all orders,
one needs to solve
the complete field equations which we have derived in the previous
section. However, to do this it is first necessary to establish
the correct relation between the fields that the string description
suggests and those appearing in the supergravity field equations.
In particular we have to find how $\widetilde b$,
$A_4$ and $C_8$ are related to $b$, $c$ and $C_0$.
It turns out that
$\widetilde b$ represents the fluctuation of the scalar $b$ of eq.(\ref{wra98})
around the background value of the ${\bf Z}_2$ orbifold \cite{bflux},
{\it i.e.} 
\be b =  \frac{1}{2}\,{(2\pi\sqrt{\alpha'})^2} +
\widetilde b~~.
\label{beta}
\ee 
The 4-form potential $A_4$ is instead the Hodge dual (in the six dimensional
sense) to the twisted scalar $c$, while the 8-form potential $C_8$ 
is the dual  (in the ten dimensional
sense) to the axion $C_0$. These duality relations, which can be
obtained from eq.s (\ref{axion45}) and (\ref{ceq}) remembering the analyticity
of $\gamma$ and $\tau$ and the absence of source terms, are
\bea
dC_8 &=& -\,{\rm e}^{2\phi}\, {}^*dC_0~~,
\label{dualc0} \\
dA_4 &=& {\rm e}^{\phi} \,{}^{*6} (d c + C_0 \,db)  - C_4 \wedge db~~.
\label{dualc}
\eea
The absence of source terms is due to the fact that the fields $c$ and $C_0$ 
are not coupled to the boundary
state of a D7-brane (see eqs. (\ref{untw9}) and (\ref{twi86})).

Using the asymptotic expressions for the various fields in these
relations, one can easily find that
\bea
C_0 &\simeq& \frac{Ng_s}{2\pi}\,\theta+...~~,
\label{C0asym} \\ 
c &\simeq& Ng_s\pi\alpha'\,\theta+...
\label{casym}
\eea
where $\theta = \tan^{-1}(x^5/x^4)$ . In the next section we are going
to determine the higher order terms and find the complete classical
solution with the asymptotic behavior described above. In particular
we will discover that the untwisted 4-form $C_4$ and the metric along
the world-volume directions of the fractional D7 brane will develop 
a non trivial profile at higher order.

\vskip 1cm
\section{The fractional D7/D3 bound state}
\label{section4}

Since the world volume of a 7-brane
is eight-dimensional, a stack of $N$ fractional D7 branes is not
immediately useful to yield information on gauge theories in four dimensions. 
To obtain a classical solution capable of describing
a four-dimensional field theory we must include also some D3-branes, 
and hence it is natural to study a bound state made of $N$ 
fractional D7-branes and $M$ fractional D3-branes. As shown in
Ref.~\cite{GRANA2}, this is a BPS configuration
which preserves eight of the sixteen supersymmetries of the type IIB
theory on the orbifold (\ref{orbifold}).  
Our task is then to solve the field equations derived in
Section \ref{section2} and 
specify the appropriate boundary action $S_{\rm b}$ for this
configuration. For the
D3 brane components, we can simply take $M$ times the action introduced in
Ref.~\cite{Bertolini:2000dk}, and thus we can write
\begin{eqnarray}
S_{~\rm b}^{\rm D3} &=& -\,M\tau_3\,
\int d^4x\,\sqrt{-\det
g_{\alpha\beta}} \left(1+\frac{1}{2\pi^2\alpha'}\, {\widetilde
b}\right)
\label{actionbound3} \\
&&+\,M\tau_3
\int C_{4}\left(1+\frac{1}{2\pi^2\alpha'}\, {\widetilde b} \right) ~+~
M\,{\widetilde \tau_3}\,
\int A_{4}
\nonumber
\end{eqnarray}
where $\tau_3$ and ${\widetilde \tau}_3$ are defined after eq.(\ref{sbuntw})
and before eq.(\ref{sbtw}).
For the D7 brane components, instead, we can take as boundary action
$N$ times the sum
of (\ref{sbuntw}) and (\ref{sbtw}). As mentioned in the previous section, the
twisted part (\ref{sbtw}) may be non-complete, but it is certainly correct
at linear order and thus yields the right 
source terms in the various field equations.
Using these ingredients and the Ansatz (\ref{met48}) and 
(\ref{f5ans}), one can show that 
eqs.(\ref{dilaeq52}) and (\ref{gammaeq28}) become
\begin{eqnarray}
\delta^{ij}\,\partial_i\partial_j \tau
&+&
{\rm i} \, 2\ky^2\,\tau_7
 \, N\,\delta(x^4)\,\delta(x^5)=0 
\label{eqtau2}~~,
\\
\delta^{ij}\,\partial_i\partial_j \gamma 
&-& {\rm i}\, 
\ky^2\,{\widetilde \tau}_3\,(2M-N)
\,\delta(x^4)\,\delta(x^5)=0 ~~.
\label{eqgamma2}
\end{eqnarray}
The holomorphic solutions to these equations can be immediately found 
and are (see also Ref.~\cite{GRANA2})
\begin{equation}
\tau =  {\rm i}\left(1 -\, \frac{Ng_s}{2\pi}\, \log \frac{z}{\epsilon}\right)
~~, 
\label{tausol}
\end{equation}
and
\begin{equation}
\gamma = \,{\rm i}\,2\pi\alpha' g_s\,\left[\frac{\pi}{g_s} + 
(2M-N)\,\log \frac{z}{\epsilon}\right]
\label{gammasol}
\end{equation}
where we have chosen the integration constants to enforce the
appropriate background values. Written in terms of 
the real supergravity fields, 
eqs.(\ref{tausol}) and (\ref{gammasol}) become
\bea
{\rm e}^\phi &=& \frac{1}{1-\frac{N g_s}{2 \pi} \,\log \frac{\rho}{\epsilon}}
~~,\label{dilsol} \\
C_0 &=& \frac{Ng_s}{2\pi}\,\theta ~~,\label{C0sol}\\
b &=& \frac{(2 \pi \sqrt{\alpha'})^2}{2}\; \frac{1+\frac{(2M-N)g_s}{\pi} \,
\log \frac{\rho}{\epsilon}}
{1-\frac{N g_s}{2 \pi} \,\log \frac{\rho}{\epsilon}} ~~,
\label{bsol}\\
&&\nonumber \\
c &=& -(2\pi\alpha') \,\theta \,g_s\,\left(2M - \frac{N}{2}\;\frac{1-\frac{2M 
g_s}{\pi} \,\log \frac{\rho}{\epsilon}}
{1-\frac{Ng_s}{2\pi} \,\log \frac{\rho}{\epsilon}} \right) ~~.
\label{csol} 
\eea
Notice that for $N=0$ this solution reduces to the one of
pure fractional D3-branes
discussed in Ref.s~\cite{Bertolini:2000dk,Polchinski:2000mx}.
It is interesting to observe, on the other hand, that putting $M=0$
and expanding in powers of $Ng_s$, we recover at first order
the solution (\ref{dila}), (\ref{C0asym}) and (\ref{casym}) obtained from
the boundary state approach. In this respect, we observe that the axion
$C_0$ does not receive corrections to higher orders while the 
twisted fields $b$ and $c$ acquire an infinite tail of logarithmic 
terms. This is to be contrasted with the solution of the
pure fractional D3 branes \cite{Bertolini:2000dk,Polchinski:2000mx}
where the twisted scalars had, instead, only terms at first order.
Thus, if one 
wants to determine
the classical profile of the twisted scalars using the boundary state 
formalism in the presence of fractional D7 branes, 
it is not sufficient to consider contributions with just 
one boundary, but it is necessary to sum over 
all contributions with an arbitrary number of boundaries as explained 
in Ref.~\cite{nbps}, which, 
due to the open/closed string duality, is equivalent 
to sum over an arbitrary number of open-string loops.  

Finally, if we insert the above expressions for $b$ and $c$ 
into eq.(\ref{Heq19}), we obtain the following
equation for the warp factor $H$:
\bea
\left(\delta^{ij}\partial_i \partial_j  + {\rm e}^{-\phi}\, 
\delta^{\ell m}\partial_\ell \partial_m\right) H  \!\!&+&\!\! 
2\ky^2\,\tau_3\,M
\,\delta(x^4)\cdots\delta(x^9)
\label{Heq54} \\
&&\hskip -2.6cm
+~(2 \pi \alpha' g_s)^2 
\frac{(2M - \frac{N}{2})^2}{\rho^2 ( 1 - 
\frac{Ng_s}{2 \pi}\log \frac{\rho}{\epsilon})^3}\,\delta(x^6)\cdots
\delta(x^9) =0 ~~.
\nonumber
\eea
In general, it is not possible to find an explicit solution of this
equation in terms of elementary functions. For $N=0$ this equation
was solved exactly in Ref.~\cite{Bertolini:2000dk}, whereas for $N=4M$,
{\it i.e.} when the last term vanishes, this equation becomes of the
same form that was considered in Ref.~\cite{Aharony:1998xz}. 
It is also interesting to observe that eq.(\ref{Heq54}) remains
non-trivial even for $M=0$. This fact means that for a system made of only 
D7 branes on orbifold
both the longitudinal metric and the 4-form $C_4$ are not trivial,
contrarily to what happens for D7 branes in flat space.
However, it should be realized that these fields start developing
only at the second order 
in the string coupling constant, as is clear from the structure 
of eq.(\ref{Heq54}) for $M=0$.

Using the full solution (\ref{dilsol})-(\ref{csol}) in the duality relations
(\ref{dualc0}) and (\ref{dualc}), and recalling, apart irrelevant additional
terms, that
\be
C_4 = \left(H^{-1} -1\right) dx^0\wedge\cdots\wedge dx^3~~,
\label{c44}
\ee
we can easily obtain the complete expressions for the 8-form $C_8$ and for the
twisted 4-form $A_4$ which are more natural from a stringy
perspective. After some algebra, we find
\be
C_8 = \frac{ \frac{Ng_s}{2 \pi}\log 
\frac{\rho}{\epsilon}}{1 - \frac{N g_s}{2 \pi} 
\log \frac{\rho}{\epsilon}} ~
d x^0 \wedge \cdots\wedge d x^7 = \left({\rm e}^\phi -1\right)
d x^0 \wedge \cdots\wedge d x^7 ~~,
\label{c8sol}
\ee
and
\be  
A_4 = \frac{(2 \pi \sqrt{\alpha '})^2}{2} ~
\frac{ \frac{(4M -N)g_s}{2 \pi}\log 
\frac{\rho}{\epsilon}}{1 - \frac{N g_s}{2 \pi} 
\log \frac{\rho}{\epsilon}} ~
d x^0 \wedge \cdots\wedge d x^3  = {\widetilde b}~
d x^0 \wedge \cdots\wedge d x^3 ~~.
\label{a4sol}
\ee

Having the complete solution, we can verify the no-force condition
and check the structure of the world-volume action of the bound state. 
If we substitute our solution into the D3-brane component 
(\ref{actionbound3}) of the boundary action,
we find, as expected, that all terms depending on the transverse coordinates
cancel, leaving a constant result. Doing the same thing
for the D7-brane components (\ref{sbuntw}) and (\ref{sbtw}), we see that
in the twisted part not all terms cancel, thus indicating the presence 
of a non-zero force. 
However, this result is unacceptable in view of the BPS properties of 
our solution.  This problem is easily overcome if we recall that
the boundary action (\ref{sbtw}) is actually justified 
only at the linear level, and thus may be non-complete. 
To construct the full boundary action we can start from
the standard expansion of the WZ part of the action for a
D7 brane, namely
\be
S_{\rm WZ} = \tau_7
\int
\left({\widehat C}_8 + {\widehat C}_6\wedge B_2
+\frac{1}{2}~{\widehat C}_4\wedge B_2 \wedge B_2\right)
+ ...
\label{swz}
\ee
where the ellipses stand for curvature terms. We now decompose the forms
${\widehat C}_8$, ${\widehat C}_6$ and ${\widehat C}_4$ into untwisted
components (denoted by $C$) and into twisted components 
along the 2-form $\omega_2$ (denoted by $A$).
For the case under consideration, the relevant expressions are
\bea
{\widehat C}_8 &=& C_8 + x\,(2\pi\sqrt{\alpha'})^2\,A_4\wedge \omega_2
\wedge \omega_2~~,\nonumber
\\
{\widehat C}_6 &=& y\, A_4\wedge \omega_2~~,
\label{chat}\\
{\widehat C}_4 &=& C_4 \nonumber
\eea
where $x$ and $y$ are numerical coefficients which will be
determined later. If we substitute eq.(\ref{chat}) into the action
(\ref{swz}) and recall that $B_2=b\,\omega_2$ with $b$
given by eq.(\ref{beta}), after some simple
manipulations we get
\bea
S_{WZ}  &=&   \tau_7
\int C_8
-
\,\frac{{\widetilde \tau}_3}{4}
\int
A_4\left[\,(2x+y) + \frac{y}{2\pi^2\alpha'}\,{\widetilde b}\,\right]
\nonumber \\
&-&
\frac{{\widetilde \tau}_3}{4}
\int C_4\,{\widetilde b}\,\left(1+\frac{\widetilde b}{4\pi^2\alpha'}\right)~~.
\label{swz1}
\eea
Notice that in writing the last expression we have used the fact that
the (understood) curvature contribution exactly cancels the term
linear in $C_4$. This fact, shown in Ref.~\cite{GRANA2}, is 
consistent with the boundary state of a fractional
D7 brane which indeed does not couple to $C_4$. Instead,
it couples to the twisted 4-form $A_4$, and matching the
corresponding charge with the boundary state result (see eq.(\ref{sbtw})) fixes
\be
2x+y=1~~.
\ee
If we substitute the classical solution
(\ref{c8sol})-(\ref{a4sol}) in eq.(\ref{swz1})
and require no force, we can see that to cancel the contribution of $C_8$ 
we must add to the boundary action the expected DBI term
\be
-\tau_7 \, \int d^8x~{\rm e}^\phi\,\sqrt{-\det
g_{ab}}~~,
\ee
while to cancel the contribution of $C_4$ we must add a term like
\be
\,\frac{{\widetilde \tau}_3}{4}
\,\int d^4x~\sqrt{-\det
g_{\alpha\beta}}~{\widetilde b}\,
\left(1+\frac{\widetilde b}{4\pi^2\alpha'}\right)
\ee
and fix $y=1/2$.
In this way the no-force condition is fully satisfied, as it should
be.
We thus conclude that the world-volume action of a fractional D7
brane consists of an untwisted part given by eq.(\ref{sbuntw}) and a
twisted part given by
\bea
\left.S_{~\rm b}^{\rm D7}\right|_{\rm T}
&=&\,\frac{{\widetilde \tau}_3}{4}
\,\int d^4x~\sqrt{-\det
g_{\alpha\beta}}~{\widetilde b}\,
\left(1+\frac{\widetilde b}{4\pi^2\alpha'}\right)\,-\,
\frac{{\widetilde \tau}_3}{4}\,\int A_4
\,\left(1+\frac{\widetilde b}{4\pi^2\alpha'}\right)\nonumber \\
&&-~
\frac{{\widetilde \tau}_3}{4}
\int C_4\,\,
{\widetilde b}\,\left(1+\frac{\widetilde b}{4\pi^2\alpha'}\right)~~.
\label{sbtwfin}
\eea
It would be interesting to confirm the structure of this boundary action
with geometrical considerations and also with explicit 
calculations of closed string scattering amplitudes on a disk with boundary 
conditions appropriate for the fractional D7 brane, similarly to what has 
been done in Ref.~\cite{MERLATTI} for the fractional D3 branes.

\vskip 1cm
\section{The probe action and the ${\cal N}=2$ gauge theory}
\label{section5}
 
The supergravity solution found in the previous section can
provide non-trivial information on its dual four-dimensional gauge
theory. To see this we use the probe technique (for a review
see Ref.~\cite{joh}) and consider a probe fractional D3-brane
carrying a gauge field $F_{\alpha\beta}$ and slowly moving in the supergravity
background produced by $M$ D3 and $N$ D7 fractional
branes. We then fix the static gauge 
and study the world-volume action of the probe, regarding
the transverse coordinates as Higgs fields $\Phi^i= (2\pi\alpha')^{-1} x^i$,
and expanding up to quadratic terms in derivatives. 
{F}rom a gauge theory point of view, 
the resulting action describes the $SU(M)\times U(1)$ 
Coulomb phase of a $SU(M+1)$ gauge theory in which the symmetry 
breaking corresponds to taking one of the D3 branes (the probe) away from the 
others at a distance $\rho=|z|$ related to the energy scale where the theory
is defined.

Applying this technique to our case, we find that the
action of a probe fractional D3-brane can be written as
\begin{equation}
S = S_0 + S_{\rm gauge}
\label{bound34}
\end{equation} 
where $S_0$ is given by eq.(\ref{actionbound3}) with $M=1$ and
\be
S_{\rm gauge} = -\frac{1}{8\pi g_s} \int d^4 x \sqrt{-\det G_{\alpha\beta}} 
~\left\{\, \frac{1}{4} \,{\rm e}^{-\phi}\,G^{\alpha \gamma} G^{\beta \delta}
F_{\alpha \beta} 
F_{\gamma \delta}
\right. 
\label{bound78}
\ee
\[
+\,\left.\frac{1}{2} \,G_{ij} G^{\alpha \beta} \partial_{\alpha} \Phi^i 
\partial_{\beta} \Phi^j \,\right\}
\left( 1 + \frac{{\tilde{b}}}{2 \pi^2 \alpha '}\right)+ \frac{1}{8\pi g_s} 
\int d^4 x\, \frac{1}{4}\,F_{\alpha \beta} {\widetilde{F}}^{\alpha \beta}
\,\left(\frac{c + C_0\,b }{2 \pi^2 \alpha'} \right)~~.
\]
where $\widetilde{F}^{\alpha \beta}=(1/2)\, 
\epsilon^{\alpha \beta\gamma\delta} F_{\gamma\delta}$.
Inserting in $S_0$ the solution for the closed string fields 
obtained in the previous section ({\it i.e.}
eq.s (\ref{tausol})-(\ref{gammasol}) 
and the Ansatz (\ref{met48})-(\ref{f5ans})), we see that 
$S_0$ becomes independent 
of the distance between the probe and 
the source branes that yield the classical solution. 
This is in agreement with the fact that there is no
interaction between a fractional D3-brane and a system of fractional
D3/D7-branes. 
 
Considering now eq.(\ref{bound78}), we see that
the dependence on the function $H$ drops out 
in this case too,
while the kinetic terms for the gauge field strength $F_{\alpha\beta}$
and the  scalar fields 
$\Phi^i$ have the same coefficient, in agreement with the fact that the 
gauge theory living on the brane has  ${\cal{N}}=2$ supersymmetry. Indeed 
one gets
\begin{equation}
S_{gauge} = -\,\frac{1}{g_{\rm YM}^{2} (\mu)}
 \int d^4 x  \left\{ \frac{1}{2} \partial_{\alpha} \Phi^i
\partial^{\alpha} \Phi^i + \frac{1}{4} F_{\alpha \beta} F^{\alpha \beta}
 \right\} 
+  \frac{\theta_{\rm YM}}{32 \pi^2} \int d^4 x F_{\alpha \beta} 
{\tilde{F}}^{\alpha \beta} 
\label{bound53}
\end{equation} 
where
\begin{eqnarray}
\frac{1}{g_{\rm YM}^{2} (\mu) } &=& \frac{1}{g^2_{\rm YM}} + 
\frac{2M-N}{8 \pi^2} \log \mu  \hspace{.5cm}; \hspace{.5cm}
g^{2}_{\rm YM} = 8\pi g_s    
\label{runn23} \\
\nonumber \\
\theta_{\rm YM} &=& (2M-N) \; \theta 
\label{runn24}
\end{eqnarray}
are the effective Yang-Mills gauge coupling and $\theta$-angle, respectively.  
The renormalization group scale is defined by
$\mu\equiv |z|/\epsilon$, while $g^{2}_{\rm YM}$ 
is the bare coupling, {\it i.e.} the value of the gauge coupling at 
the ultraviolet cutoff $\mu=1$. 

Eq.(\ref{runn23}) clearly shows that 
$g_{\rm YM} (\mu)$ is the running coupling constant of
an ${\cal{N}}=2$ supersymmetric gauge theory 
with gauge group $SU(M)$ and $N$ hypermultiplets in the fundamental 
representation. 
This is precisely the field theory 
living on the system of $M$ D3-branes and $N$ D7-branes, where 
the gauge vector multiplet corresponds to open strings stretched 
between two fractional D3-branes, 
while the hypermultiplets correspond to strings stretched 
between the D3 and the D7-branes. The reason why the hypermultiplet
kinetic term is absent in $S_{gauge}$ 
is just because the probe is a D3-brane only, and therefore 
there are no 3-7 strings that can give rise to massless fields on the 
probe world-volume. Of course, this theory is ultraviolet 
free only for $N \le 2M$

{F}rom eq.(\ref{actionbound3}) one sees that on the geometric locus defined by
\begin{equation}
|z_e| \;=\; \epsilon \;{\rm e}^{- \pi\,/\,(2M-N)\,g_s}~~,
\label{enh}
\end{equation}
the D3-brane probe becomes tensionless, thus indicating the presence of 
an enhan\c{c}on. At distances smaller 
than $|z_e|$ the probe has negative tension, while at the 
enhan\c{c}on extra light degrees 
of freedom come into play~\cite{Polchinski:2000mx}. This means 
that the supergravity approximation leading to the solution described 
in section 4 is not valid anymore, and that the region 
of space-time $\rho < |z_e|$ is excised. Notice that the 
vanishing of the tension of the probe at the 
enhan\c{c}on is consistent with the fact that fractional branes are 
tension-full because of the presence 
of a non-vanishing $B_{(2)}$ flux, $b$ \cite{bflux}. Indeed, by 
using eq.(\ref{enh}), one can write $\gamma$ as follows
\begin{equation}
\gamma = 2\pi \,i\, \alpha'\,g_s \left(2M-N\right)\,\log z/|z_e| ~~,
\label{gammaen}
\end{equation}
and see that the quantity
${\rm Im}\,\gamma \equiv {\rm e}^{-\phi}\,b$, which is proportional
to the probe tension, vanishes
at the enhan\c{c}on, since there the
fluctuation of the $b$ field cancels precisely its background value. 

{F}rom eq.(\ref{runn23}) one can immediately recognize what is the 
meaning of the enhan\c{c}on from the gauge 
theory point of view. This is the scale where the gauge coupling diverges 
(which in QCD is called $\Lambda_{\rm QCD}$) 
and where non-perturbative corrections become relevant. 
This means that, as discussed in the Introduction, 
the supergravity solution is only able to reproduce the perturbative 
moduli space of the gauge theory, 
while the appearance of the enhan\c{c}on prevents from using the 
classical solution to analyze
the strong-coupling properties of the gauge theory. The translational 
dictionary between supergravity and 
gauge quantities can then be summarized as follows
\begin{eqnarray}
\frac{4\pi}{g_{\rm YM}^2(\mu)} = \frac{1}{(2\pi\sqrt{\alpha'})^2}\; 
e^{-\phi} \, b  &,& 
\frac{\theta_{\rm YM}}{2\pi} = - \frac{1}{(2\pi\sqrt{\alpha'})^2} 
\,\frac{1}{g_s}\, (c + C_0 \, b) \label{gt} ~~,\\
\nonumber \\
\Lambda_{\rm UV} = (2\pi\alpha')^{-1}\, \epsilon &,& \Lambda_{\rm QCD} 
= (2\pi\alpha')^{-1}\, |z_e| ~~,  
\end{eqnarray}
where in eq.(\ref{gt}) the dilaton includes also its background value. 
It is interesting to observe 
that the presence of D7-branes lowers the enhan\c{c}on radius. 
This can be seen explicitly from eq.(\ref{enh}). In 
particular, when $N=2M$ the enhan\c{c}on vanishes, the gauge 
coupling constant does not run anymore 
and the gauge theory becomes conformal, as expected for 
a supersymmetric ${\cal{N}}=2$ gauge theory with
gauge group $SU(M)$ and with $2M$ hypermultiplets transforming according to the
fundamental representation 
of $SU(M)$. Notice that also in the conformal case the contribution of the
twisted fields in eq.(\ref{Heq54}) for $H$ does not vanish making the 
solution of eq.(\ref{Heq54}) quite not trivial. 
The twisted field contribution vanishes, 
however, for $N=4M$ and in this case the field equation 
for $H$ reduces to the one discussed in 
Ref.~\cite{Aharony:1998xz}. The vanishing of the twisted contribution is a 
consequence of the fact that the coupling of a fractional D7-brane 
to the twisted fields is a factor $1/4$ smaller than that of 
a fractional D3-brane. In this case, however, the theory is not
ultraviolet free.

A distinctive feature of the D3/D7 system 
with respect to that of pure fractional 
D3-branes of Ref.~\cite{Bertolini:2000dk} is 
that the scalar fields given in eq.s (\ref{dilsol})-(\ref{csol}) 
are expressed as an infinite series in the open string coupling. 
However, the scalar field combinations which have 
a meaning at the gauge theory level, namely those appearing 
in eq.(\ref{gt}), are exact at one-loop, as 
expected for a ${\cal N}=2$ super Yang-Mills theory. This non-trivial 
cancellation is a (higher loop) check of the validity 
of the gauge/gravity correspondence. 

\vskip 1.5cm 
{\large {\bf Acknowledgments}}
\vskip 0.5cm
\noindent
We would like to thank I. Pesando for sharing with us his insights on 
various aspects 
related to this work. M.B. would also like to thank R. Russo for useful 
e-mail correspondence. 
M.B. is supported by INFN.

\appendix
\vskip 1.5cm  
\section{The boundary state description of the Dp brane}  
\label{appea}  
\renewcommand{\theequation}{A.\arabic{equation}}  
\setcounter{equation}{0}  
\noindent  
The boundary state for a D$p$ brane with $r$ directions of its 
world-volume outside and $s=p-r$ directions along the
orbifold ${\rm I\!R}^{4}/{\bf Z_2}$, can be derived by 
factorizing the one-loop vacuum amplitude of the open strings stretching 
between two such branes. This amplitude is given by
\begin{equation}  
Z = Z_{1} + Z_{g}
\label{z}
\end{equation}
where
\begin{eqnarray}
Z_{1} &=& \frac{1}{2} \int_0^\infty \frac{ds}{s}\, {\rm Tr}_{\rm NS-R}\,
\big[\, P_{\rm GSO}\, {\rm e}^{- 2 \pi s (L_0 -a)} \,\big] \nonumber
\\
&=& \frac{1}{2}\, \frac{V_{p+1}}{\left(8\pi^2\alpha'\right)^{(p+1)/2}}\,
 \int_0^\infty \frac{ds}{s^{(p+3)/2}}\, \frac{1}{2} \left[\, \frac{ f_3^8(q) - 
f_4^8(q)
- f_2^8(q)}{ f_1^8(q)}\, \right]~~,
\label {z1} \\
Z_{g} &=& \frac{1}{2}\, \int_0^\infty \frac{ds}{s}\, {\rm Tr}_{\rm NS-R}\, 
\big[\,g\,  P_{\rm GSO} \,{\rm e}^{- 2 \pi s (L_0 -a)}\, \big]  \nonumber
\\
&=& \frac{V_{r+1}}{2^s \, \left(8 \pi^2 \alpha'\right)^ {(r+1)/2}}\,
 \int_0^\infty \frac{ds}{s^{(r+3)/2}} \left[ \,
\frac{f_3^4(q)\, f_4^4(q)}{f_1^4(q)\,f_2^4(q)} - 
\frac{f_3^4(q)\,f_4^4(q)}{f_1^4(q)\,f_2^4(q)}\,
 \right] 
\label{zg}
\end{eqnarray}
where $P_{\rm GSO}$ is the GSO projection, $g$ is the orbifold parity, 
$q= {\rm e}^{- \pi s}$, the $ f$'s are the standard one-loop 
modular functions and the intercept $a$ is 1/2 [0] in the 
NS [R] sector. Notice the appearance of the important factor $2^{-s}$ in
eq.(\ref{zg}) that is due to the integration over the
bosonic zero modes along the orbifolded directions \cite{marco1}.

After performing the modular transformation 
$s \to t=1/s $, $Z_1$ and $Z_g$
can be interpreted 
as tree level closed string amplitudes between two
untwisted and two twisted boundary states respectively, that is 
\begin{eqnarray}
Z_{1} &=& \frac{\alpha' \pi}{2} \int_0^\infty dt \,\,
\,^U\!\langle {\rm D}p|\,{\rm e}^{- \pi t (L_0 + {\tilde{L}}_0 -2a )}
|{\rm D}p\rangle^U ~~,
\label{z11} \\
 Z_{g} &=& \frac{\alpha' \pi}{2} \int_0^\infty dt \,\,
\,^T\!\langle {\rm D}p|\,{\rm e}^{- \pi t ( L_0 + {\tilde{L}}_0 ) }
|{\rm D}p\rangle^T ~~.
\label{zc}
\end{eqnarray}

{F}rom eq.(\ref{z1}) it is immediate to realize that $Z_1$ is one half of the 
amplitude for D$p$-branes in flat space, and therefore the untwisted part 
of the boundary state is simply 
\begin{equation}
\label{bound1}  
\ket{{\rm D}p}^U =\,\, \frac{T_p}{2\sqrt{2}}\, \left(\,
\ket{{\rm D}p}_{\rm NS}^U \,+ \,\ket{{\rm D}p}_{\rm R}^U\, \right)
\end{equation}
where $\ket{{\rm D}p}_{\rm NS}^U$ and $\ket{{\rm D}p}_{\rm R}^U$ 
are the usual boundary states for a bulk D$p$-brane 
given in Ref.s~\cite{bs,bs1}.

{F}rom eq.(\ref{zg}) we can see that the twisted amplitude for
a fractional D$p$-brane with $s$ directions along the orbifold is the same as
the one for a fractional D$r$-brane entirely outside the orbifold,
apart from a factor $2^{-s}$. Therefore, using eq.(\ref{zc}), we can deduce
that the boundary state $\ket{{\rm D}p}^T $ is similar to the 
boundary state for a fractional D$r$-brane transverse to the orbifold,
but with an extra factor of
$2^{-s/2}$ in its normalization. In conclusion, we get
\begin{equation}
\label{boundt}  
\ket{{\rm D}p}^T \, = \,- \,\frac{1}{2^{s/2}}\,
\frac{T_r}{2\sqrt{2}\,\pi^2 \alpha' } \,
\left(\,\ket{{\rm D}p}_{{\rm NS}}^T + \ket{{\rm D}p}_{{\rm R} }^T\, \right)
\end{equation}
where
\begin{equation}  
\label{proi}  
\ket{{\rm D}p}_{{\rm NS,R}}^T
= \frac{1}{2}\,\left(\,\ket{{\rm D}p,+}_{{\rm NS,R} }^T
\,+\, \ket{{\rm D}p,-}_{{\rm NS,R} }^T\, \right)~~,
\end{equation}
and the Ishibashi states  $\ket{{\rm D}p,\eta}_{\rm NS,R}^T$ are 
\begin{equation}
\label{bound2}  
\ket{{\rm D}p,\eta}^T_{\rm NS}=   
\ket{{\rm D}p_X}^T\ket{{\rm D}p_\psi,\eta}_{\rm NS}^T  
\end{equation}  
in the NS-NS twisted sector, and   
\begin{equation}
\label{bound3}  
\ket{Dp, \eta}^T_{\rm R}=   
\ket{Dp_X}^T\ket{Dp_\psi,\eta}_{\rm R}^T  
\end{equation}
in the R-R twisted sector \footnote{In eq.s (\ref{bound2}) and (\ref{bound3})  
we omit the ghost and superghost contributions   
which are not affected by the   
orbifold projection.},  
with
\begin{eqnarray}  
|{\rm D}p_X \rangle^T  &=&\delta^{(5-r)}({\widehat q}^i-y^i)   
\prod_{n=1}^{\infty} {\rm e}^{-\frac{1}{n}  
\alpha_{-n}^\alpha  
\eta_{\alpha \beta}\tilde\alpha_{-n}^\beta}  
\prod_{n=1}^{\infty} {\rm e}^{\frac{1}{n}  
\alpha_{-n}^i\tilde\alpha_{-n}^i}  
\nonumber
 \\
& &~\times\,\prod_{r=\frac{1}{2}}^{\infty} {\rm e}^{- \frac{1}{r}  
\alpha_{-r}^\ell\tilde\alpha_{-r}^{\,\ell}}  
\prod_{\beta}
|p_\beta = 0\rangle
\prod_{{i}}
|p_i = 0\rangle~~,
\nonumber
\\  
|{\rm D}p_{\psi} , \eta \rangle_{NS}^T  &=&   
\prod_{r=\frac{1}{2}}^{\infty} 
{\rm e}^{{\rm i}\eta\psi_{-r}^\alpha \eta_{\alpha \beta}   
\tilde \psi_{-r}^\beta}   
\prod_{r=\frac{1}{2}}^{\infty} {\rm e}^{-{\rm i}\eta\psi_{-r}^i 
\tilde \psi_{-r}^i}  
\prod_{n=1}^{\infty} e^{{\rm i}\eta\psi_{-n}^\ell   
\tilde \psi_{-n}^{\,\ell}} |{\rm D}p_{\psi} , 
\eta \rangle ^{(0)\,\,T}_{\rm NS}~~,  
\nonumber 
\\  
|{\rm D}p_{\psi} , \eta \rangle_{R}^T &=&  
\prod_{n=1}^{\infty} {\rm e}^{{\rm i}\eta\psi_{-n}^\alpha 
\eta_{\alpha \beta}   
\tilde \psi_{-n}^\beta}
\prod_{n=1}^{\infty} {\rm e}^{-{\rm i}\eta\psi_{-n}^i \tilde \psi_{-n}^i}
\prod_{r=\frac{1}{2}}^{\infty} {\rm e}^{{\rm i}\eta\psi_{-r}^\ell    
\tilde \psi_{-r}^{\,\ell}} |{\rm D}p_{\psi} , 
\eta \rangle ^{(0)\,\,T}_{\rm R} ~~.
\nonumber 
\end{eqnarray}
In these expressions the longitudinal indices $\alpha, \beta$ take 
values $0,1, \ldots r$, 
the transverse index $i$ takes values  $r+1,
\ldots, 5$, while the index $\ell$ labels the orbifold directions 
(to avoid further clutter, in the above formulas we have explicitly 
considered only the case in which all these orbifold
directions are longitudinal). 
The zero-mode part of the boundary state has a non trivial structure in both  
sectors; in the NS-NS sector
it is given by~\cite{bs1} 
\begin{equation}
\label{bound7}  
|{\rm D}p_{\psi} , \eta \rangle ^{(0)\,\,T}_{\rm NS}=  
\left(\widehat C {\widehat{\gamma}}^{6} \dots {\widehat{\gamma}}^{5+s} 
\frac{1 + {\rm i}\eta\widehat\gamma}{1 + {\rm i}\eta}  
\right)_{LM}|L\rangle|\widetilde M\rangle  
\end{equation}
where $\widehat\gamma^\ell$ are the gamma matrices and $\widehat C$ 
the charge conjugation matrix of $SO(4)$, $\widehat\gamma=\widehat\gamma^{6}...
\widehat\gamma^{9}$, and, finally,
$|L\rangle$ and $|\widetilde M\rangle$ are
spinors of $SO(4)$. The matrices of $SO (4)$  satisfy the following
relations under transposition
\begin{equation}
{\widehat{C}}^{\,t} = {\widehat{C}}~~~~,~~~~
 {\widehat{\gamma}}^{\ell\,t} = {\widehat{C}} 
\,\,{\widehat{\gamma}}^{\ell} \,\, {\widehat{C}}^{-1}~~.
\label{transpo76}
\end{equation} 
In the R-R sector, instead, we have  
\begin{equation}
\label{bound8}  
|{\rm D}p_{\psi} , \eta \rangle ^{(0)\,\,T}_{\rm R}=  
\left(\bar C\bar\gamma^0...\bar\gamma^r\frac{1+ {\rm i}\eta\bar\gamma}  
{1+{\rm i}\eta}  
\right)_{AB}|A\rangle|\widetilde B\rangle  
\end{equation}
where
$\bar\gamma^\alpha$ are the gamma matrices and $\bar C$ the charge   
conjugation matrix  
of $SO(1,5)$, $\bar\gamma=\bar\gamma^0...\bar\gamma^5$,  
and, finally,  $|A\rangle$ and $|\widetilde B\rangle$ are spinors of $SO(1,5)$.
The matrices of $SO(1,5)$ satisfy the following relations under
transposition
\begin{equation}
{\bar{C}}^t = - {\bar{C}}~~~~,~~~~ {{\bar{\gamma}}^{\alpha\,t}} = - {\bar{C}} 
\,\,{\bar{\gamma}}^{\alpha} \,\, {\bar{C}}^{-1}~~.
\label{transpo84}
\end{equation}
  
In order to compute the fermionic zero-mode contribution to $Z_g$  
in eq.(\ref{zc}) it is convenient to write explicitly the
conjugate vacuum states, which are given by~\cite{bs1}
\begin{equation}
{}^{(0) T}_{\rm NS} \langle {\rm D}p_{\psi} , \eta | =
\langle \widetilde M | \langle L |
\left( \widehat C {\widehat{\gamma}}^{6} \dots {\widehat{\gamma}}^{5+s} 
\frac{1 - {\rm i}\eta\widehat\gamma}{1 - {\rm i}\eta}  
\right)_{LM}    
\label{bra36}
\end{equation}
for the twisted NS-NS sector, and 
\begin{equation}
{}^{(0) T}_{\rm R} \langle {\rm D}p_{\psi} , \eta | = \langle \widetilde B |
\langle A| 
\left(\bar C\bar\gamma^0...\bar\gamma^r\frac{1+ {\rm i}\eta\bar\gamma}  
{1- {\rm i}\eta}  \right)_{AB}     
\label{brar34}
\end{equation}
for the R-R sector. 
Using the previous expressions and performing some straightforward 
algebra, it is possible to 
show that
\begin{equation}
{}^{(0) T}_{\rm NS} \langle {\rm D}p_{\psi} , \eta_1 | 
 {\rm D}p_{\psi} , \eta_2 \rangle ^{(0)\,\,T}_{\rm NS} 
= 4 \delta_{\eta_1 \eta_2;1}
\label{sandns39}
\end{equation}
for the NS-NS sector, and
\begin{equation}
{}^{(0) T}_{\rm R} \langle {\rm D}p_{\psi} , \eta_1 | 
 {\rm D}p_{\psi} , \eta_2 \rangle ^{(0)\,\,T}_{\rm R} = - 4 \delta_{\eta_1 \eta_2;1}
\label{sandr28}
\end{equation}
for the R-R sector.

Finally, by saturating the boundary states described above with
the untwisted closed string states explicitly given in 
Ref.s~\cite{antonella,bs,bs1}, 
one obtains eq.s (\ref{untw9}), while by saturating
the twisted components with the twisted states given in Ref.~\cite{anto}
one gets eq.s (\ref{twi86}).


\end{document}